\documentclass[3p,times]{elsarticle}
 
\usepackage{ecrc}
 
\volume{00}
 
\firstpage{1}
 
\journalname{Journal of Computational Physics}
 
\runauth{Alvarado et al.}
 
\jid{jcp}
 
\jnltitlelogo{J. Comp. Phys.}

\usepackage{lineno,hyperref} 
\modulolinenumbers[5] 
 
\usepackage{mathtools}
\usepackage[usenames, dvipsnames]{xcolor}
\usepackage{subcaption}

\usepackage{tikz}
\usepackage{tikz-qtree}
\usepackage{tkz-berge}
\usetikzlibrary{intersections}
\usepackage{tkz-euclide}
\usetkzobj{all}
\usepackage{sansmath}
\usepackage{bm}
\usetikzlibrary{shadings,intersections}
\usetikzlibrary{shapes.geometric, arrows.meta}
\usepackage{tikz-3dplot}
\usetikzlibrary{calc,3d,decorations.markings,backgrounds,positioning,intersections,shapes}
\def\radius{1.mm} 

\usepackage{sansmath}
\usepackage{nicefrac}
\usepackage{caption}
\usepackage{subcaption}
\usepackage{graphicx}
 
\usepackage[]{hyperref}
\hypersetup{
    colorlinks,%
    citecolor=blue,%
    filecolor=black,%
    linkcolor=blue,%
    urlcolor=black
}
 
\usepackage{cleveref}
\usepackage{tabularx}
\usepackage{multicol}

\usepackage{float}
\usetikzlibrary{shapes.geometric, arrows.meta}
\tikzstyle{startstop} = [rectangle, rounded corners, minimum width=3cm, minimum height=1cm, text width=3cm, text centered, draw=black]
\tikzstyle{io} = [trapezium, trapezium left angle=80, trapezium right angle=100, minimum width=2cm, minimum height=1cm, text width=3cm, text centered, draw=black]
\tikzstyle{process} = [rectangle, minimum width=3cm, minimum height=1cm, text width=3cm, text centered, draw=black]

\tikzstyle{decision} = [diamond, minimum width=1cm, minimum height=1cm, text width=0.5cm, text centered, draw=black]
\tikzstyle{arrow} = [thick,->,>=stealth]

\biboptions{sort&compress}
 
\usepackage{scrextend}
\usepackage[bold]{hhtensor}
 
\begin{document}
 
\begin{frontmatter} 
\title{Monte Carlo Raytracing Method for Calculating Secondary Electron Emission from Micro-Architected Surfaces} 


\author[UCLA_MSE]{Andrew Alvarado}
\ead{obispo23@ucla.edu}
\author[UCLA_MSE]{Hsing-Yin Chang}
\author[UCLA_MAE]{Warren Nadvornick}
\author[UCLA_MAE]{Nasr Ghoniem}
\author[UCLA_MSE,UCLA_MAE]{and Jaime Marian}
 
\address[UCLA_MSE]{Department of Materials Science and Engineering, University of California, Los Angeles, CA 90095, USA}
\address[UCLA_MAE]{Department of Mechanical and Aerospace Engineering, University of California, Los Angeles, CA 90095, USA}
 
\begin{abstract}
Secondary electron emission (SEE) from inner linings of plasma chambers in electric thrusters for space propulsion can have a disruptive effect on device performance and efficiency. SEE is typically calculated using elastic and inelastic electron scattering theory by way of Monte Carlo simulations of independent electron trajectories. However, in practice the method can only be applied for ideally smooth surfaces and thin films, not representative of real material surfaces. Recently, micro-architected surfaces with nanometric features have been proposed to mitigate SEE and ion-induced erosion in plasma-exposed thruster linings. 
In this paper, we propose an approach for calculating secondary electron yields from surfaces with arbitrarily-complex geometries using an extension of the \emph{ray tracing} Monte Carlo (RTMC) technique. We study nanofoam structures with varying porosities as representative micro-architected surfaces, and use RTMC to generate primary electron trajectories and track secondary electrons until their escape from the outer surface. Actual surfaces are represented as a discrete finite element meshes obtained from X-ray tomography images of tungsten nanofoams. At the local level, primary rays impinging into surface elements produce daughter rays of secondary electrons whose number, energies and angular characteristics are set by pre-calculated tables of SEE yields and energies from ideally-flat surfaces. We find that these micro-architected geometries can reduce SEE by up to 50\% with respect to flat surfaces depending on porosity and primary electron energy.\\
\end{abstract}
 

\end{frontmatter} 


\section{Introduction}\label{intro}

The release of electrons from a material surface exposed to a primary electron beam, known as secondary electron emission (SEE), is an important phenomenon with applications in a wide variety of physical processes, such as in electron multiplication devices \cite{bruining2016physics, shih1997secondary}, electron microscopes \cite{seiler1983secondary, reimer2013scanning}, and plasma devices \cite{dunaevsky2003secondary, raitses2006measurements, phelps1999cold, zhurin1999plasma, stangeby2000plasma, boyle2004electrostatic, kaganovich2007kinetic, sydorenko2008plasma}, among others. While SEE can be induced to amplify electron currents, such as during photoemission spectroscopy \cite{willis1974secondary}, it can also be detrimental for performance, such as in the case of the \emph{multipactor} effect in radio frequency devices \cite{kishek1998multipactor}. 

In the case of Hall thrusters for electric propulsion \cite{zhurin1999plasma}, an electrostatic sheath forms between the plasma and the inner lining of the plasma-facing surface material. This sheath potential acts as a thermal insulator and as a deterrent of current flow that protects the wall from particle discharges \cite{hobbs1967heat}. SEE weakens this sheath potential \cite{hobbs1967heat, kaganovich2007kinetic,sydorenko2008plasma,ahedo2004influence}, which has detrimental effects for the stability of the thruster, as is known to occur as well in magnetic fusion devices and radiofrequency plasma sources \cite{stangeby2000plasma,kaganovich2007kinetic}. Thus, as a crucial phenomenon affecting the efficiency of these devices, there is an increasing interest in mitigating --or at least controlling-- secondary electron emission. A direct method to reduce the overall SEE yield is to engineer the structure of the material surface, leading to a class of surfaces known under the umbrella term of \emph{microarchitected} surfaces.

Aside from early efforts in surface texture development to control SEE \cite{curren1986carbon,baglin2000secondary}, the use of advanced characterization and new processing techniques to develop microarchitected surfaces and experimentally determine the reduction in SEE yield is relatively recent \cite{patino2016secondary,huerta2018secondary,swanson2018modeling}.
There are now numerous examples of successful designs that are seen to lower the secondary electron yield \cite{raitses2006operation,patino2016secondary,ye2013suppression,yang2015nanostructured}. These surfaces can be fabricated {\it ex situ} and deposited over existing chamber walls to achieve the desired level of functionalization.

The theory of secondary electron emission is generally well known and has been studied for decades \cite{tagkey1962iii,dekker1952theory,bruining2016physics}. However, studies of how the surface geometry and morphology affect the rate of electron emission are relatively limited. Modeling and simulation can play an important role in predicting the expected reduction rates of SEE before a costly effort of surface texture development, fabrication, and testing needs to be mounted. Simulations involving surfaces with grooves \cite{pivi2008sharp}, `velvet'-like fibers \cite{huerta2018secondary} and open-cell structures \cite{swanson2018modeling} have been recently carried out, showcasing the versatility of numerical simulation but also its relatively high computational cost. In this paper, we present a two-pronged simulation approach in which SEE yields and energy spectra are precomputed for ideally flat surfaces, and are later used to describe the constitutive response at the local material point level of a discretized surface with arbitrary geometry. The connection between both descriptions is made via a ray-tracing Monte Carlo algorithm coupled to an intersection detection algorithm. The method simulates individual electron tracks, one at a time, and generates secondary tracks on the basis of incident energies and angles sampled from the precomputed physical relations. The number of tracks that escapes the surface is tallied and compared to the total number of simulated tracks to compute the effective secondary electron emission yield.

The paper is structured as follows: in Sec.~\ref{model} we provide a description of the intersection detection algorithm and the raytracing Monte Carlo method. In Sec.~\ref{fem} we describe the discretized model of the microarchitected surface considered, while in Sec.\ \ref{results} we demonstrate the validity of the method by reproducing results for flat surfaces and open cell foam unit cells, followed by SEE yield calculations in in actual foam structures of various porosities. We conclude with a brief discussion section and the conclusions. 

\section{Computational model}
\label{mod}
Our model is based on a raytracing model to track particle trajectories from a random point above the surface as they are directed towards the material. This \emph{primary} ray is defined by its energy $E$ and angle of incidence with respect to a laboratory (global) frame of reference. Once this primary ray is generated, the next step is to determine whether its trajectory intersects the material surface, discretized into a finite element mesh, i.e.~whether the the ray crosses a surface element of the mesh. The algorithm used to detect such intersections is an extesnion of the M\"{o}ller and Trumbore algorithm \cite{moller2005fast}, which we briefly review in the following.


\subsection{Intersection detection algorithm}
\label{model}
Next, we provide a brief overview of the M\"{o}ller-Trumbore (M-T) method \cite{moller2005fast}. 
The procedure is described for triangular surface elements, although it can be extended to other geometric shapes in a straightforward manner.
A ray defined by an origin $\vec{O}$ and a direction $\vec{D}$ is defined by the equation:
\begin{equation}
\vec{R}(t) = \vec{O} + t \vec{D}
\label{MT1}
\end{equation}
where $t$ is a scaling parameter that defines the length of the ray. If the vertices of the surface element triangle $\vec{V}_0,\vec{V}_1,\vec{V}_2$ are known, any point on the element can be defined by
\begin{equation}
\vec{T}(u,v) = (1 - u - v)\vec{V}_0 + u\vec{V}_1 + v\vec{V}_2.
\label{MT2}
\end{equation}
Where ${u}$ and ${v}$ are barycentric coordinates that define the plane of a triangle, and satisfy $u \geq 0$, $v \geq 0$, and $ u^2 + {v}^2 \leq 1$. Note that the surface element normal can be determined as\footnote{Assuming that the vertices are given a counter-clockwise order as seen from a direction opposing the normal.}:
$$\vec{n} = \frac{\vec{s}}{\|\vec{s}\|},~\vec{s}\equiv\left(\vec{V_1}-\vec{V_0}\right)\times\left(\vec{V_2}-\vec{V_0}\right)$$
As outlined by M\"{o}ller and Trumbore, the intersection point can be uniquely found by equating eqs.~\eqref{MT1} and \eqref{MT2} above: 
\begin{equation} 
\vec{O} + t \vec{D} = (1 - u - v)\vec{V}_0 + u \vec{V}_1 + v\vec{V}_2
\end{equation}
Rewriting and rearranging in terms of matrices gives
\begin{equation}
\begin{bmatrix}
-\vec{D}, & \vec{V}_1 - \vec{V}_0, & \vec{V}_2 - \vec{V}_0 \\
\end{bmatrix}
\begin{bmatrix}
t \\
u \\
v \\
\end{bmatrix} 
= \vec{O} - \vec{V}_0
\end{equation}
Defining $\vec{L}_1 = \vec{V}_1 - \vec{V}_0,$ $\vec{L}_2 = \vec{V}_2 - \vec{V}_0$, and $ \vec{T} = \vec{O} - \vec{V}_0$, the solution can be obtained through Cramer's rule:
\begin{equation}
\begin{bmatrix}
t\\
u\\
v\\
\end{bmatrix}  
= \frac{1} {\left(\vec{D}\times\vec{L}_2\right)\cdot \vec{L}_2} 
\begin{bmatrix}
\left(\vec{T}\times\vec{L}_1\right)\cdot \vec{L}_2 \\
\left(\vec{D}\times\vec{L}_2\right)\cdot \vec{T} \\
\left(\vec{T}\times\vec{L}_1\right)\cdot \vec{D}
\end{bmatrix}
\end{equation} 
The algorithm must scan through all elements of the structure, following for each element the steps above in search for possible intersections. For large meshes, the computation can rapidly become prohibitive. For this reason, several checks are employed to quickly determine if an intersected point lies on a triangular element:
\begin{enumerate}
\item First, a back-face culling technique is applied so that if the normal of a triangular element is in the direction of an incoming ray, that element is ignored.
\item Second, if the ray is parallel to the plane of a triangle within an allowed tolerance, that element is ignored. 
\item If the above two checks are satisfied, then a last check is made to determine the conditions for barycentric parameters $u$ and $v$. If all conditions are satisfied, then the algorithm determines the intersection point within a triangular element. 
\item Finally, a micro-architected surface may have many surface elements intersecting a given ray, only the closest is considered.
\end{enumerate}

\subsection{Generation of secondary rays}
\label{iris}
Once a collision is detected via the M-T procedure, the incidence angle of the primary ray on the selected surface element is determined as:
$$\alpha=\cos^{-1}\left(\frac{\vec{D}\cdot\vec{n}}{\|\vec{D}\| ~ \|\vec{n}\|}\right)$$
Note that this angle of incidence is a local variable (given in a relative frame of reference). The pair $(E,\alpha)$ is then used to sample from bivariate relations giving, first, the number of secondary rays per incident primary ray \cite{chang2017calculation}:
\begin{equation}
\begin{split}
\gamma(E, \alpha) = 3.05 + 1.7949\times10^{-5} \alpha^2 + 6.15 \times10^{-7}E^2 
- \frac{371.317 - \exp{\left(0.04826\alpha\right)}}  {87.602+E}  +\\
- 1.974 \times10^{-3}E - 0.1097\cos{(0.1201+2.469\alpha)}
\end{split}
\label{seey}
\end{equation}
Specifically, this function gives the SEE yield from a flat tungsten surface for a primary electron beam of energy $E$ at an incident angle $\alpha$. So once a collision has been confirmed, we evaluate this function with the primary ray's energy and angle of incidence and the resulting fractional yield is rounded up or down using a uniform random number generator\footnote{For example, if a yield $\gamma=1.2$ is obtained, we sample uniformly in the interval $(0,1]$ and if the random number $\xi_1\ge0.2$, then we round the yield down to $\gamma=1$. Else, it is rounded up to 2.}. This means that rays that produce yields $\gamma<1$ may result in no secondary electron emission. If one secondary ray is emitted, the energy of the resulting electron is obtained by evaluating the accompanying function \cite{chang2017calculation}:
\begin{equation}
E_{\rm SEE} (E, \alpha) = 0.195 E + 	1.6\times 10^{-4} E^2 + 0.148 \alpha \sin{(E)}
 + 3.44\times10^{-15}E \alpha^7 - 6.54~~{\rm [eV]}
\end{equation}
In the event that more than one secondary rays are produced, the energy of one of them is obtained as:
\begin{equation}
E_1 = E_{\rm SEE} \sqrt{\xi_2}.
\end{equation}
and emitted with an angle sampled from a cosine distribution:
$$\beta_1=\sin^{-1}\left(\xi_2^{1/2}\right)$$
The second emitted ray then has an outgoing angle of
$$\beta_2 = \frac{\pi}{2} - \beta_1$$
and energy
$$E_2  = \cos^2{\beta_1} E_{\rm SEE}$$
If $E_1,E_2<E_{c}$, the corresponding ray is terminated, where $E_{c}$ is a threshold energy equal to $E_{c} = \Phi + E_{f}$ with $\Phi$ the material's workfunction and $E_{f}$ the material's Fermi energy, both in eV \cite{streitwolf1959theorie,kotera1989monte}. Therefore, once non-primary rays reach an energy less than the threshold, they no longer have the ability to emit secondary electrons. 

The rays are tracked one at a time, from intersection to intersection, after they terminate. Each primary ray crates a tree. For example, a primary ray that generates two secondary rays will add two branches to the tree. 
$$\lambda_p \rightarrow \lambda'_{d} + \lambda''_{d}$$ $$\lambda_p \lambda'_{d}  \lambda''_{d}$$ 
Once the calculations of a given branch are completed the algorithm moves to the branch. If a given branch spawns another branch (third `generation'), then it is added to the end of the tree and the tree moves to next branchß.
$$\lambda'_d \rightarrow \lambda'_{gd}$$ $$\lambda_p \lambda'_{d} \lambda''_{d} \lambda'_{gd}.$$  This sequence repeats until a tree has no further branches and encounters a null element.  With this procedure, each ray's energy, angle, generation, and starting point/termination is tracked. Periodic boundary conditions are used along the $x$ and $y$ directions, while a flat boundary is used at $z=z_{\rm min}$ to mimic a solid substrate beneath the foam.  
A ray (of any generation) that is found to reach $z_{\rm max}$ with an polar angle between $\pm\nicefrac{pi}{2}$ is considered and counted as a secondary electron emission event.

\begin{figure}[h]
\centering
\fbox{
\tikzstyle{startstop} = [rectangle, rounded corners, minimum width=2cm, minimum height=1.5cm,text centered, font=\sffamily, draw=black, fill=red!30]
\tikzstyle{input} = [trapezium, trapezium left angle=70, trapezium right angle=110, minimum width=1cm, minimum height=1.2cm, text centered, font=\sffamily, draw=black, fill=blue!20]
\tikzstyle{process} = [rectangle, minimum width=3cm, minimum height=5cm, text centered, font=\sffamily, draw=black, fill=orange!30]
\tikzstyle{decision} = [diamond, minimum width=3cm, minimum height=1.5cm, text centered, font=\sffamily, draw=black, fill=green!30]
\tikzstyle{arrow} = [thick,->,>=stealth]
\centering
\begin{tikzpicture}[scale=0.75, node distance=1.5cm,
    every node/.style={fill=white, font=\sffamily, transform shape}, align=center]
\node (kernel)	at (0,0)  [process] {\LARGE \bf{Calculation Kernel} \\ \\ \\ \\ \Large M-T intersection algorithm \\ \Large {$\downarrow$}\\ \Large {$\alpha$} \\ \\ \Large Sampling functions \\ \Large {$\downarrow$}\\ \\ \Large {$\gamma(N), E$}};
\node (primary)	[input, above of=kernel, yshift=4cm]    {\large Generate primary rays \\ \LARGE {$(x_{p},y_{p},z_{p}), E_{p}$}};   
\node (fixed)	[above of=primary, xshift=-1.25cm,yshift=2cm] {\Large Fixed \\ \Large incidence \\ \Large angle};
\node (random)	[above of=primary, xshift=1.25cm,yshift=2cm] {\Large Random \\ \Large incidence \\ \Large angle};                
\node (mesh)	[process, left of=kernel, xshift=-5.25cm]    {\huge {\bf Geometry} \\ \\ \\ \\ \huge \bm{$V_{0}, V_{1}, V_{2}$} \\ \\ \huge \bm{$n$} \\ \\};
\node (decision_0)	[decision, below of=kernel, yshift=-3cm]    {\Large {$i \leq N$}};  
\node (ray_i)	[startstop, below of=decision_0, yshift=-1.5cm]    {\Large Ray 1 \\ \Large Ray 2 \\ \Large \bm{$\vdots$} \\ \Large Ray $i$};    
\node (decision_1)	[decision, right of=kernel, xshift=3cm, yshift=1.75cm]    {\Large {$z_{i} > z_{\rm max}$}};    
\node (decision_2)	[decision, right of=kernel, xshift=3cm, yshift=-1.75cm]    {\Large {$E_{i} > E_{c}$}};    
\node (yield)	[startstop, right of=decision_1, xshift=2.5cm]    {\Huge $\gamma_{\rm eff}$};
\node (terminate)	[startstop, right of=decision_2, xshift=2.5cm]    {\large Terminate \\ \large ray $i$};

\draw [arrow] (primary) -- (kernel);
\draw [arrow] (fixed) -- (fixed.south|-primary.north);
\draw [arrow] (random) -- (random.south|-primary.north);
\draw [arrow] (decision_1) -- node[above,pos=0.5] {\large Yes} (yield);
\draw [arrow] (decision_2) -- node[above,pos=0.5] {\large No} (terminate);
\draw [arrow] (decision_2) -- node[right,pos=0.5] {\large Yes} (decision_1);
\draw [arrow] (decision_1) |- node[right,pos=0.125] {\large No} node[pos=1] {\Large {$(x_{i},y_{i},z_{i}), E_{i}$}} ([xshift=1.5cm,yshift=1cm]kernel.north) -| ([xshift=1.5cm]kernel);
\draw [arrow] (ray_i) -| (decision_2);
\draw [arrow] (kernel) -- (decision_0);
\draw [arrow,name path=line 1] (decision_0) -| node[above,pos=0.125] {\large No} ([xshift=-1cm]kernel.west) |- (primary);
\path [name path=line 2] (mesh) -- (kernel);
\path [name intersections={of = line 1 and line 2}];
\coordinate (S)  at (intersection-1);
\path [name path=circle] (S) circle(\radius);
\path [name intersections={of = circle and line 2}];
\coordinate (I1)  at (intersection-1);
\coordinate (I2)  at (intersection-2);
\draw[ thick,->,>=stealth] (I1) -- (kernel);
\draw [thick] (I2) -- (mesh);
\tkzDrawArc[color=black,thick](S,I1)(I2);
\draw [arrow] (decision_0) -- node[right,pos=0.5] {\large Yes} (ray_i);

\end{tikzpicture}
}
\caption{Flowchart of the raytracing Monte Carlo code.\label{flow}}
\end{figure}
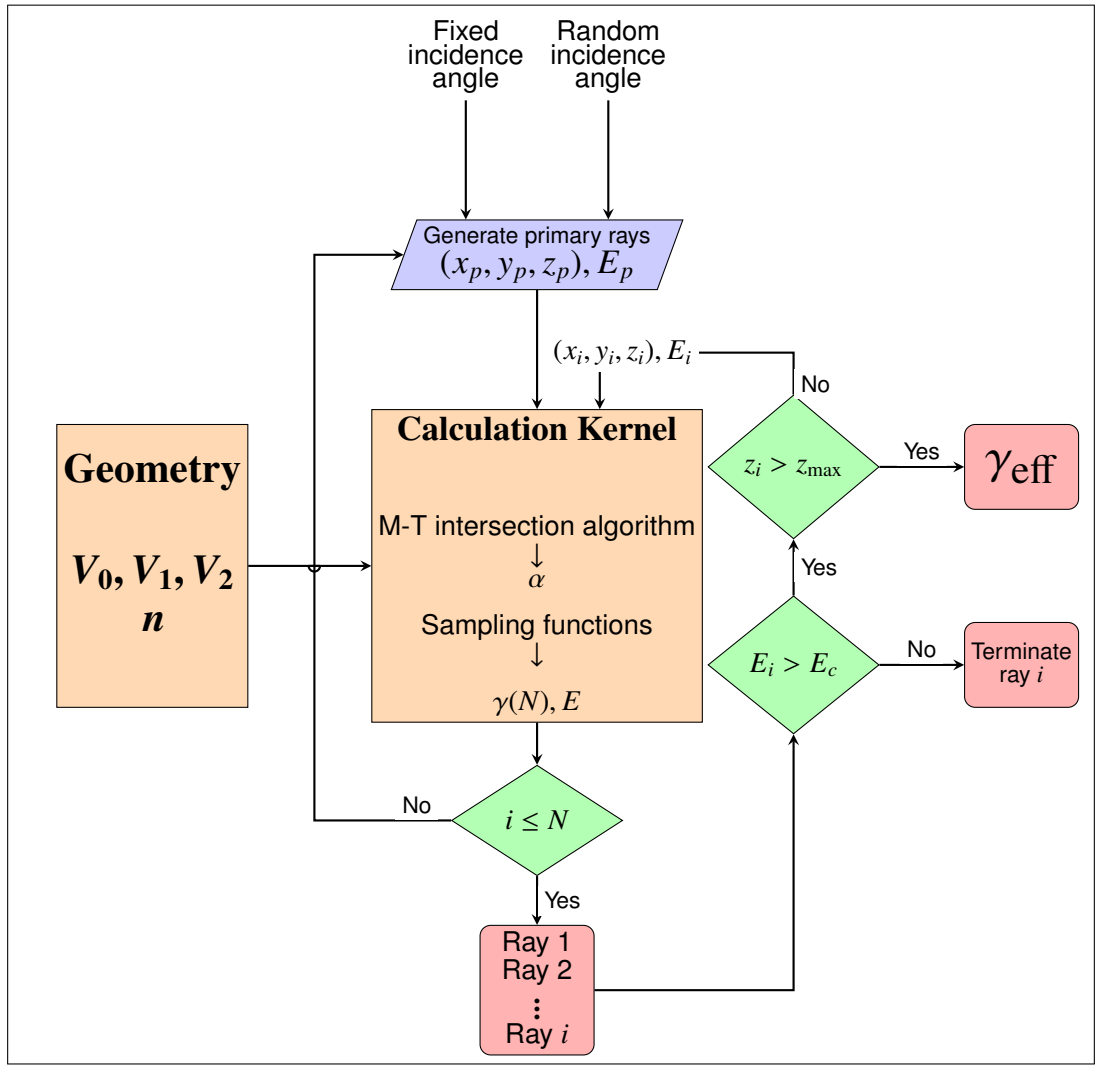

This raytracing approach is trivially parallelizable, in the sense that each primary ray is independent and the geometry of the foam remains unaltered for all rays. This allows the method to run on multiple replicas and quickly generate large subsets of data, allowing the control of number of initial rays, energy, position and direction. 
The flow diagram of the entire process is given in Figure \ref{flow}.

\subsection{Finite element model and Surface geometry development}\label{fem} 

The foam model is computationally reconstructed from a series of grayscale X-ray tomography images of a real foam with 65 pores per inch (PPI) and approximately 4\% volume fraction $V_f$ \cite{GAO2018319}. Each image is filtered and stacked to create a three-dimensional array with elements assigned either a value of 0 (space) or 1 (material). This voxel representation of the foam can be manipulated to change the foam's morphology, allowing for SEE yield comparisons to be drawn between foams of various porosities. The volume fraction of the foam is increased by adding layers of material voxels to the surface voxels, which are identified by their immediate proximity to voxels of value equal to zero. To prevent the growth of large, flat surfaces for the higher volume fraction foams which can adversely affect meshing quality, some randomness is included in the voxel layering process. This procedure is used to generate foams with $V_f=$ 4\%, 6\%, 8\% and 10\%. Volume fractions are computationally determined by summing the material voxels and dividing by the total number of voxels in the domain. Although the procedure to generate these surfaces is general, this particular structure is based on foams with pore and ligament sizes of approximately 270 and 80 $\mu$m, respectively \cite{GAO2018319}. 

\begin{figure}[h]
    \centering
    \begin{subfigure}[t]{0.5\textwidth}
        \centering
        \includegraphics[width=\columnwidth]{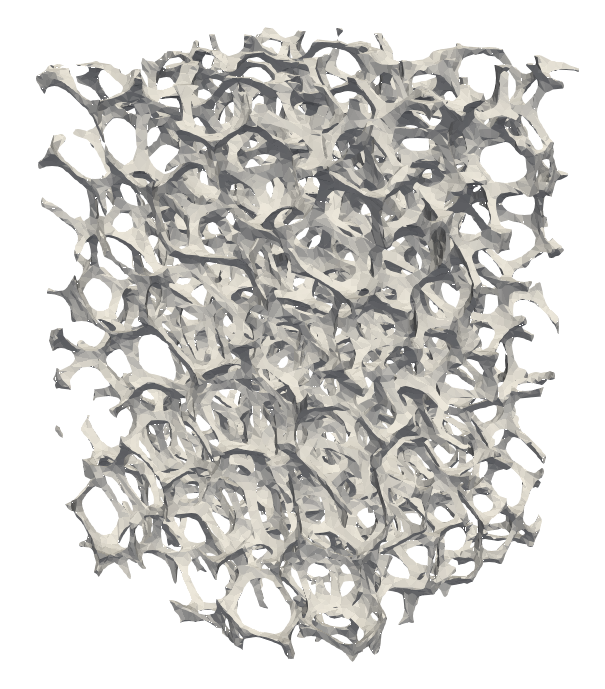}
        \caption{\label{mesh}}
    \end{subfigure}%
    \begin{subfigure}[t]{0.5\textwidth}
        \centering
        \includegraphics[width=\columnwidth]{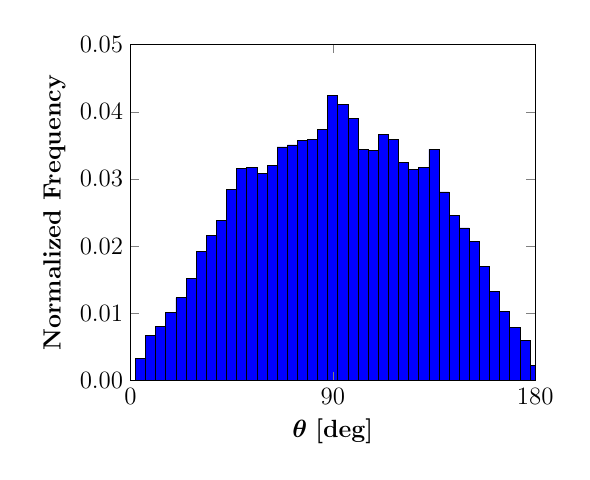}
        \caption{\label{normals}}
    \end{subfigure}
    \caption{(a) Finite element model of a real micro-architected foam structure rendered from X-ray tomography images. (b) Histogram of surface element normals.}
    \label{ashby}
\end{figure}

An iso-surface routine is run on the voxel model to create the finite element model used for the raytracing study. The number of triangular surface elements generated is typically of order $10^7$ and is reduced using a mesh coarsening routine to order $10^5$. This ensures reasonable simulation runtimes and introduces a wider variety of element angles in the finalized mesh. An example of a finite element foam model is shown in Figure \ref{mesh}. An important aspect of the mesh is the distribution of surface element normals that will be encountered by the simulated rays. Figure \ref{normals} shows a histogram of surface element normals for the geometry shown in Fig.\ \ref{mesh}. As the figure reveals, the distribution of normals is not uniform, with a maximum observed for orientations near 90$^{\circ}$ (perpendicular to the $z$-axis). This is indicative of (nonuniform) pore shapes elongated along the vertical axis, a factor which will be invoked in Sec.\ \ref{res:arch} to explain some particular results.

\section{Results and Discussion}\label{results}
Next we show several key results of the method. First, we carry out a series of verification tests to confirm the correctness of the implementation. We then deploy the verified methodology to cases of practical interests such as the nanofoam architected surface.  

\subsection{Verification}
\begin{figure}[h]
\centering
\includegraphics[width = 0.9\textwidth, angle = 0]{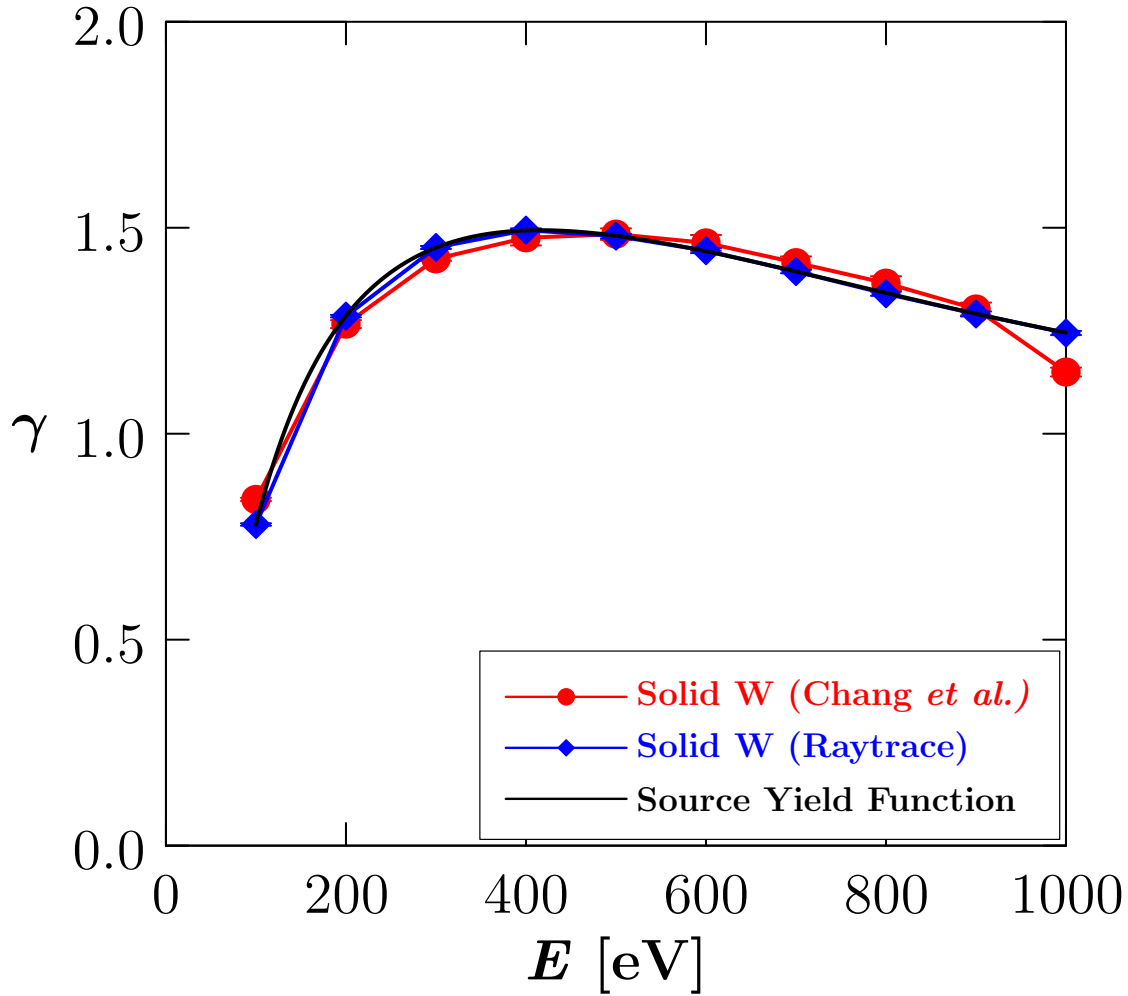}
\caption{SEE yield as a function of primary energy for normal incidence on ideally flat W surfaces obtained using (i) scattering Monte Carlo (raw data from ref.\ \cite{chang2017calculation}), (ii) sampling functions given in eq.\ \eqref{seey}, and (iii) using the raytracing model described here.}
\label{test1}
\end{figure}

The first test performed involves studying flat W surfaces by sampling from the SEE yield surface as a function of incident energy and angle calculated in our previous work \cite{chang2017calculation} (eq.\ \eqref{seey}). This simply verifies the implementation of the sampling procedure. The thickness of the sample is chosen so as to ensure that the no primary ray has enough energy to traverse it up to 1000 eV.  $10^5$ rays per energy point are simulated, distributed over 10 computational cores. As Figure \ref{test1} shows the results using the raytracing method match exactly those given by the sampling function for normal incidence. The actual data from scattering Monte Carlo from which the sampling function is obtained is also shown for reference.


\begin{figure}[h]
    \centering
    \begin{subfigure}[t]{0.5\textwidth}
        \centering
        \includegraphics[width=\columnwidth]{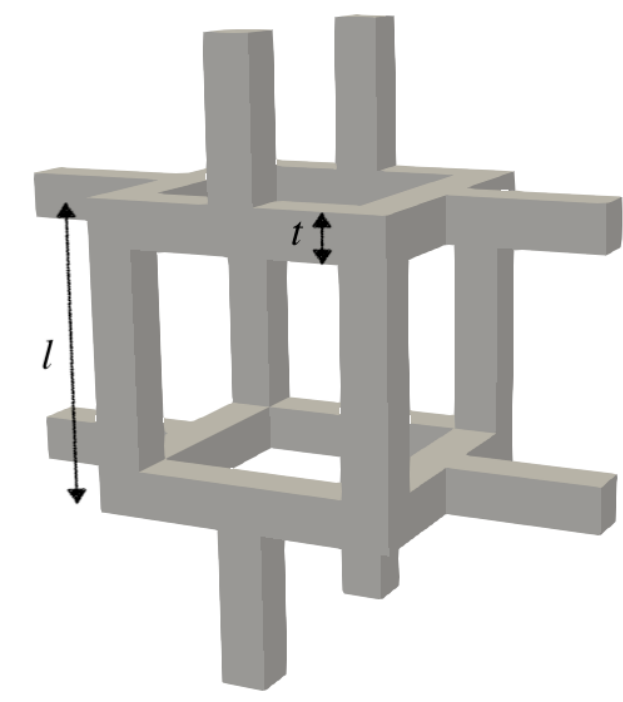}
        \caption{\label{fig:figure2a}}
    \end{subfigure}%
    \begin{subfigure}[t]{0.5\textwidth}
        \centering
        \includegraphics[width=\columnwidth]{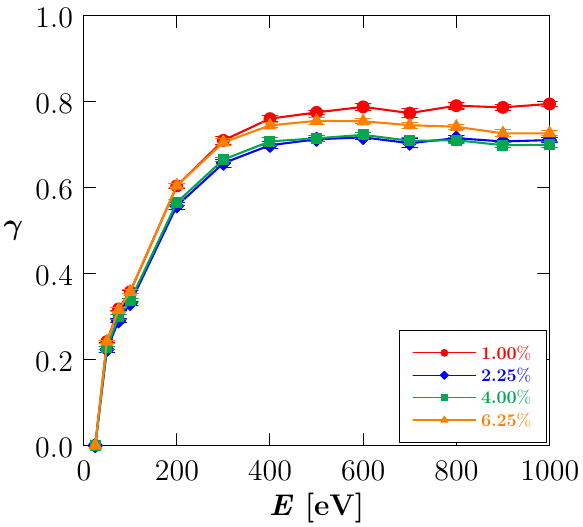}
        \caption{\label{fig:figure2b}}
    \end{subfigure}
    \caption{(a) Image of a cubic open cell foam structure with the cage size $l$ and ligament size $t$ indicated. (b) Secondary electron yield versus electron beam energy at 0 degree incidence from the cubic cage.}
    \label{ashby}
\end{figure}
The second test is performed on an open cell (also known as `open cage') structure, shown in Figure \ref{fig:figure2a}, which is a simple way to represent foams with arbitrary porosity. It was proposed by Gibson and Ashby, who express the solid volume fraction of the structure as \cite{gibson1999cellular}:
\begin{equation}
v_f=\frac{\rho_c}{\rho} =C \left(\frac{t}{l}\right)^2
\end{equation}
where $\rho_c$ is the relative density, $\rho$ is the density of the material of which the cell is made, $t$ is the ligament thickness, $l$ is the cell size, and $C$ is a proportionality constant with a value around 28. These cells can be used as repeat units of periodic arrangements simulating foams of arbitrary size. 
Next we calculate the SEE yield for an open cage with volume fractions of 1.00, 2.25, 4.00, and 6.25\%. A full-dense flat surface is assumed to lie beneath the cage, as to simulate an underlying solid substrate. The results are shown in Figure \ref{fig:figure2b} for normal primary electron incidence. Interestingly, the maximum yield is obtained for the lowest volume fraction. This is an artifact due to the simulation setup, as in that case the ligaments are too thin to absorb any primary electrons and most of the primary rays hit the bottom substrate and secondary electrons are able to escape unimpeded.

%

\subsection{Micro-Architectured Foam Structures}\label{res:arch}
\begin{figure}[h]
\centering
\includegraphics[width = 0.85\textwidth, angle = 0]{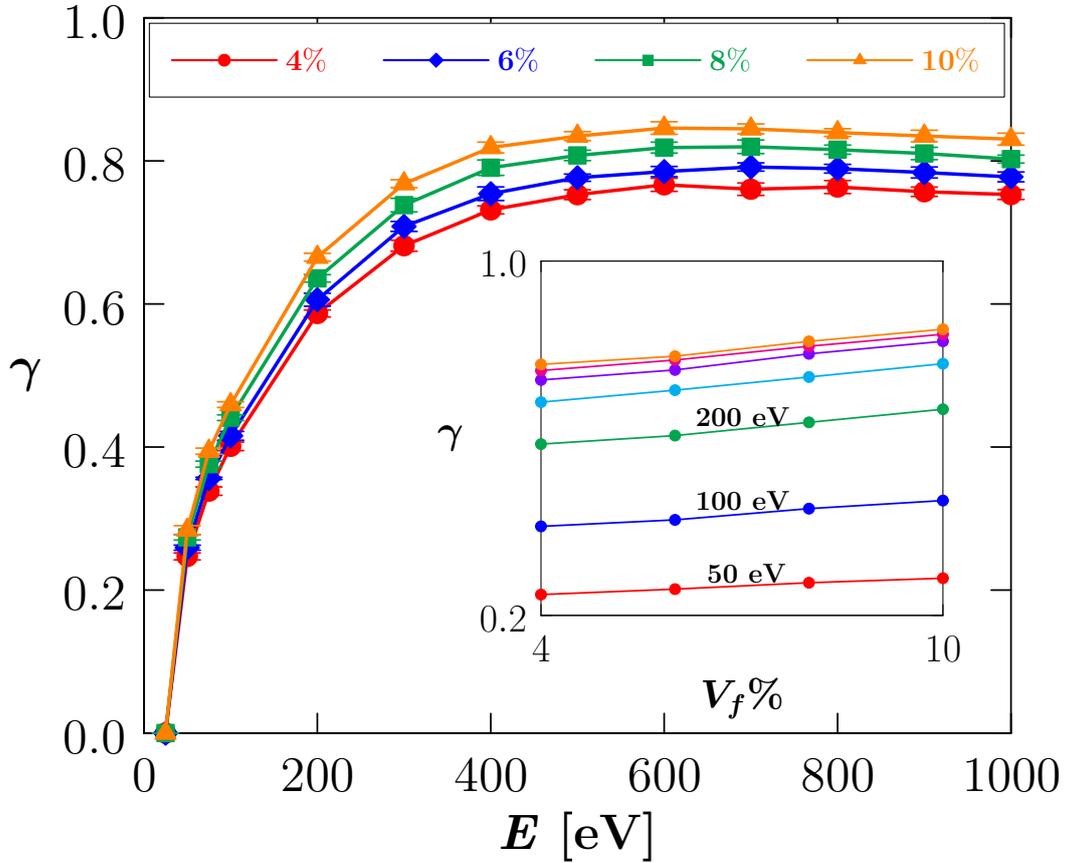}
\caption{SEE yield versus electron beam energy initially projected from 0 degree incidence to the foam at varying volume-fraction percentages. The inset shows (in increasing order) the dependence of the yield with volume fraction for primary energies equal to 50, 100, 200, 300, 400, 500, and 600 eV.}
\label{see1}
\end{figure}

Four foam structures as the one shown in Figure \ref{mesh} with varying volume fractions are are studied. Figure \ref{see1} shows the secondary electron emission yield for solid volume fractions of 4, 6, 8, and 10\% in the 50-to-1000-eV energy range using normal incidence.
The inset to the figure shows the dependence of the yield with the material volume fraction at energies of 50, 100, 200, 300, 400, 500, and 600 eV. In the high porosity range explored here the dependence of the SEE yield on $V_f$ is clearly linear in the high porosity range explored here. For the sake of comparison, the maximum SEE yield for $V_f=4\%$ (which occurs for $E=600$ eV) is approximately 0.76, compared with a value of 1.49 for the flat surface (from Fig.\ \ref{test1}). This decrease in SEE yield by about a factor of two is indicative of the potential performance gains that micro-architected surfaces might offer relative to fully dense surfaces. It is also worth noting that the results in Fig.\ \ref{fig:figure2b} for an open cell cage with 4\% material volume fraction is approximately 0.71, suggesting that such a simple model could be an acceptable surrogate for more complex geometries.
\begin{figure}[h]
\centering
\includegraphics[width = 0.85\textwidth, angle = 0]{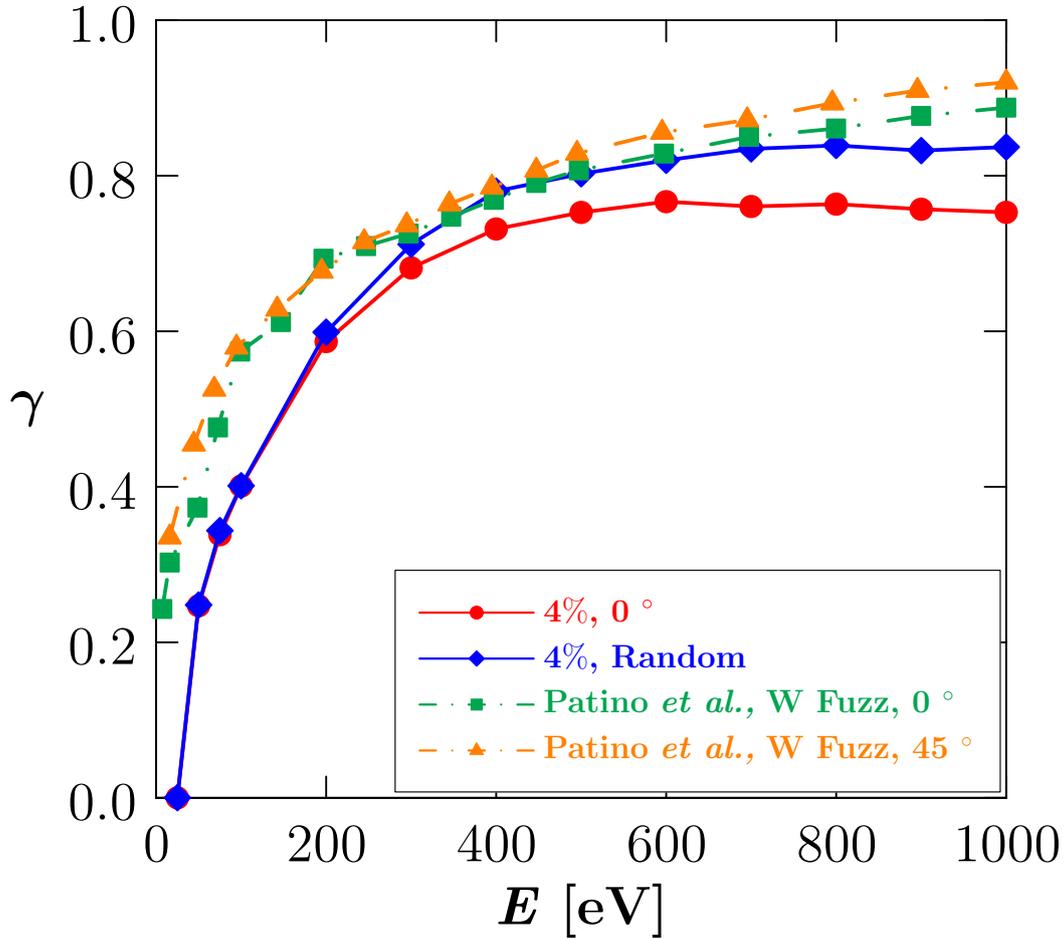}
\caption{SEE yield versus electron beam energy for normal and random incidence primary electrons. Experimental results by Patino {\it et al.}~\cite{patino2016secondary} are shown for comparison.}
\label{fuzz}
\end{figure}

\begin{figure} [h]
\centering
\includegraphics[width = 0.9\textwidth, angle = 0]{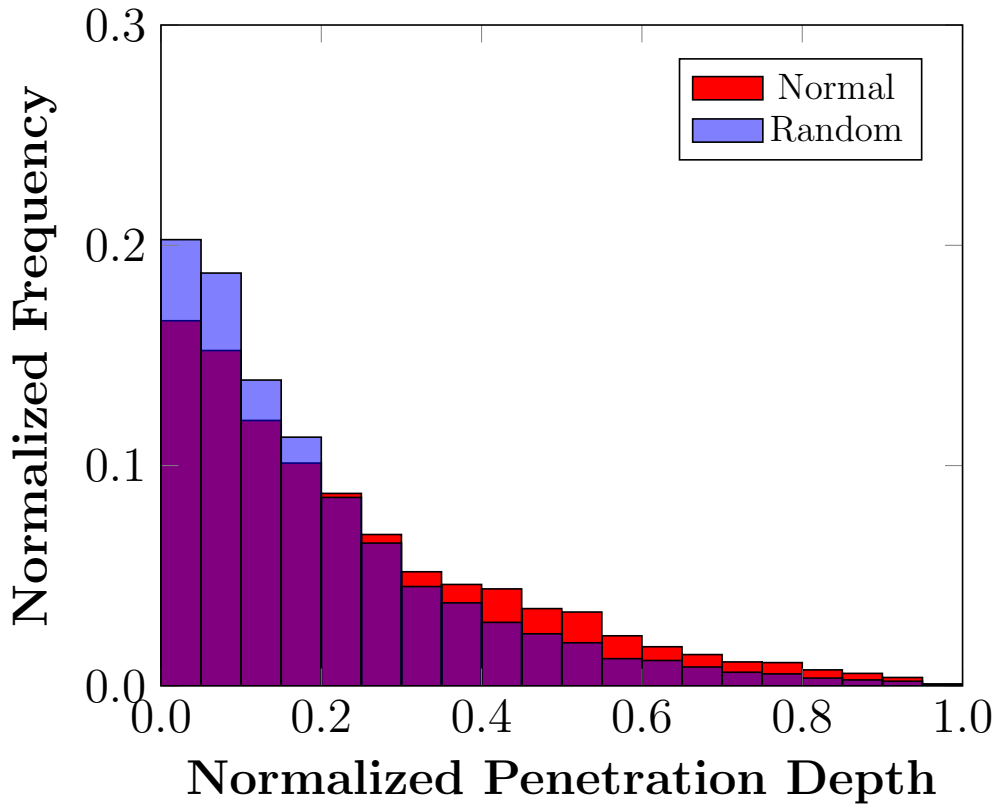}
\caption{Ppenetration depth of electrons in a 4\% volume fraction foam with an electron beam energy of 100 eV at normal and random primary electron incidence.}
\label{pen}
\end{figure}

Next we study the effect of the incident angle on the results by carrying out a study considering random primary incidence vs.~just normal incidence. Given that the distribution of surface normals in the foam is not uniform (cf.| Fig.\ \ref{normals}), we do expect some differences between both cases. As Figure \ref{fuzz} shows, these differences are more pronounced at high incident energies where primary rays with random incidence are able to impinge on high-angle surface elements at shallower depths and create more secondary electrons than normal rays, which are capable of penetrating deeper distances and become absorbed there. This is the reason behind the higher yields for random incidence shown in the figure. To confirm this, one can compute the penetration profile of the electron beam in both cases, obtained by tallying the depths at which electrons --regardless of what `generation' they belong to-- thermalize and become absorbed by the material. The results are shown in Figure \ref{pen}, where it can be seen that, overall, normal-incidence rays penetrate further than the random incidence counterparts. The normalized depth can be scaled to typical foam thicknesses of approximately three millimeters, as described by Gao {\it et al.} \cite{GAO2018319}.

Finally, we compare our results with experimental data. While no measurements have been made for W micro-foams, there are data available on He-plasma-exposed W surfaces \cite{patino2016secondary} , which are seen to develop a nano-tendril structure (commonly known as `fuzz') at temperatures above approximately 900$^\circ$ \cite{BALDWIN2009886}. These fuzz structures with characteristic ligament sizes of 10$\sim$20 nm resemble open foam surfaces with high porosity and can therefore be considered for comparison against our raytracing Monte Carlo calculations. The results are also shown in Fig.~\ref{fuzz}, where very good agreement is found between the calculations for the foam with $V_f=4\%$ and the fuzz surfaces used in the experiments. The good match between the experimental data for 0 and 45$^\circ$ primary incidence serves as indirect indication that the distribution of surface normals in the fuzz is closer to uniform. 

%

\section{Discussion and conclusions}\label{conc}

Secondary electron emission is an important process in materials exposed to charged particle distributions (cf.\ Sec.\ \ref{intro}). Measurements of SEE yields are challenging, and modeling and simulation can play an instrumental part in predicting the response of complex surfaces to electron exposure. Models can also be used as a way to pre-assess the suitability of a specific surface morphology prior to developing costly fabrication techniques \cite{lukkassen2003advanced,garcia2016commercial,singh2018survey}. For this, computational methods must display sufficient efficiency to parse through the parametric space, which can be large if one takes into account the multiple length scales of the problem, such as pore size, ligament size, total thickness, total exposed area, etc. Here, we develop an experimentally-validated methodology that simulates electron irradiation on a surface using rays generated randomly with a given set of properties. This technique, known as `raytracing' Monte Carlo, is routinely used in the visual graphics industry to create shades and lighting effects \cite{appel1968some,whitted1979improved,chalmers2002practical}. The interaction of each ray with the surface consists of a mathematical determination of the possible intersections with it, and a physical description of an incident ray impinging on a surface. Per se, our model is trivially-scalable and can run on multiple processors without any communication cost.

However, to be fully applicable, the methods proposed in his paper must satisfy two premises: (i) that a surface structure with arbitrary geometric complexity can be reduced by discretization to a piecewise collection of flat surface elements on which to apply the secondary electron physics of flat surfaces, and (ii) that electron irradiation on these discretized geometries can be effectively simulated with individual rays representing electron trajectories. In this sense, the raytracing approach proposed here is not conceptually too different from the original neutron transport Monte Carlo methods developed several decades ago to study neutronics in nuclear reactors \cite{pelowitz2005mcnpxtm,wasastjerna2008using}. The M-T algorithm acts as the bridge between the discretized material surface and the raytracing approach. A feature not to be overlooked is the pre-computation of SEE yields and energy distributions for flat (ideal) surfaces. It is in these calculations where all the physics around electron-matter interactions is contained, such that the problem of SEE from complex surfaces can be easily separated into a `physics' part (in idealized scenarios) and a `geometry' part (discretized to take advantage of the physical calculations). This division affords a great deal of versatility, so that as long as a sufficiently fine mesh can be generated one can conceivably study geometric features as fine as nanopillars, internal voids, surface islands, or even asperities associated with surface roughness.

This is the case in this work, where porous foams with volume fractions $<10\%$ have been studied. Foams of this type, with thicknesses as thin as just a few microns lying on solid substrates are shown to reduce the SEE yield by over 50\%. This is an encouraging finding to promote the use of these micro-architected structures in materials exposed to charged plasmas. Work to extend this methodology to dielectric materials of interest in plasma thrusters for electric propulsion (BN \cite{meezan2001linear,satonik2012modification,satonik2014effects,levchenko2018recent}) is currently underway.

To conclude, we list the main findings of our work:
\begin{enumerate}
\item We have developed a raytracing Monte Carlo approach that generates random electron trajectories and determines their intersection with solid surfaces via the M\"oller-Trombore algorithm.
\item Each intersection is characterized by an impinging angle and energy, from which partial electron emission trajectories can be generated from pre-calculated relations. 
\item All trajectories are tracked until either an electron emission is recorded or until the energy of the ray falls below the threshold energy for escape.
\item We find that micro/nano foams with 96 to 90\% porosity reduce the net SEE yield by approximately 50\% in W surfaces.
\item We find very good agreement between our full approach and measurements of SEE yields in W-fuzz surfaces across a wide energy range.
\end{enumerate}

\section*{Acknowledgements}
The authors acknowledge support from the Air Force Office of Scientific Research (AFOSR), through award number FA9550-11-1-0282 with UCLA.


\end{document}